\documentclass[sigconf, authorversion]{acmart}

\usepackage[export]{adjustbox}
\usepackage{enumitem}
\usepackage{listings}
\usepackage{subfig}

\copyrightyear{2021}
\acmYear{2021}
\setcopyright{rightsretained}
\acmConference[SIGCSE '21]{Proceedings of the 52nd ACM Technical Symposium on Computer Science Education}{March 13--20, 2021}{Virtual Event, USA}
\acmBooktitle{Proceedings of the 52nd ACM Technical Symposium on Computer Science Education (SIGCSE '21), March 13--20, 2021, Virtual Event, USA}
\acmDOI{10.1145/3408877.3432580}
\acmISBN{978-1-4503-8062-1/21/03}

\begin{document}
\fancyhead{}

\title{Nifty Web Apps}
\subtitle{Build a Web App for Any Text-Based Programming Assignment}

\author{Kevin Lin, Sumant Guha, Joe Spaniac, Andy Zheng}
\orcid{0000-0001-9946-3635}
\affiliation{%
  \department{Paul G. Allen School of Computer Science \& Engineering}
  \institution{University of Washington}
  \streetaddress{185 E Stevens Way NE}
  \city{Seattle}
  \state{Washington}
  \country{USA}
  \postcode{98195}
}
\email{{kevinl, guhas2, jspaniac, succion}@cs.uw.edu}

\begin{abstract}
While many students now interact with web apps across a variety of smart devices, the vast majority of our Nifty Assignments \cite{nifty} still present traditional user interfaces such as console input/output and desktop GUI. In this tutorial session, participants will learn to build simple web apps for programming assignments that execute student-written code to dynamically respond to user interactions resulting in a more modern app experience. Our approach requires up to 75\% less code than similar desktop GUI apps while requiring few (if any) modifications to existing assignments. Instructors and students alike can run and modify these web apps on their own computers or deploy their apps online for access from any smart device at no cost. The tutorial presents examples from CS1 and CS2 courses in Python and Java, but the ideas apply generally.
\end{abstract}



\newsavebox{\console}
\begin{lrbox}{\console}%
\begin{minipage}[t]{2.8in}%
\begin{lstlisting}[aboveskip=0pt, belowskip=0pt, columns=fullflexible, basicstyle=\linespread{1.2}\ttfamily]
$ javac Autocomplete.java
$ java Autocomplete
Query: Sea
608660 Seattle, Washington, United States
 33025 Seaside, California, United States
 26909 SeaTac, Washington, United States
 24168 Seal Beach, California, United States
 22858 Searcy, Arkansas, United States

Query:
\end{lstlisting}%
\end{minipage}%
\end{lrbox}%

\begin{teaserfigure}
  \subfloat{\usebox{\console}}%
  \hfill%
  \subfloat{\includegraphics[height=2.15in, valign=t]{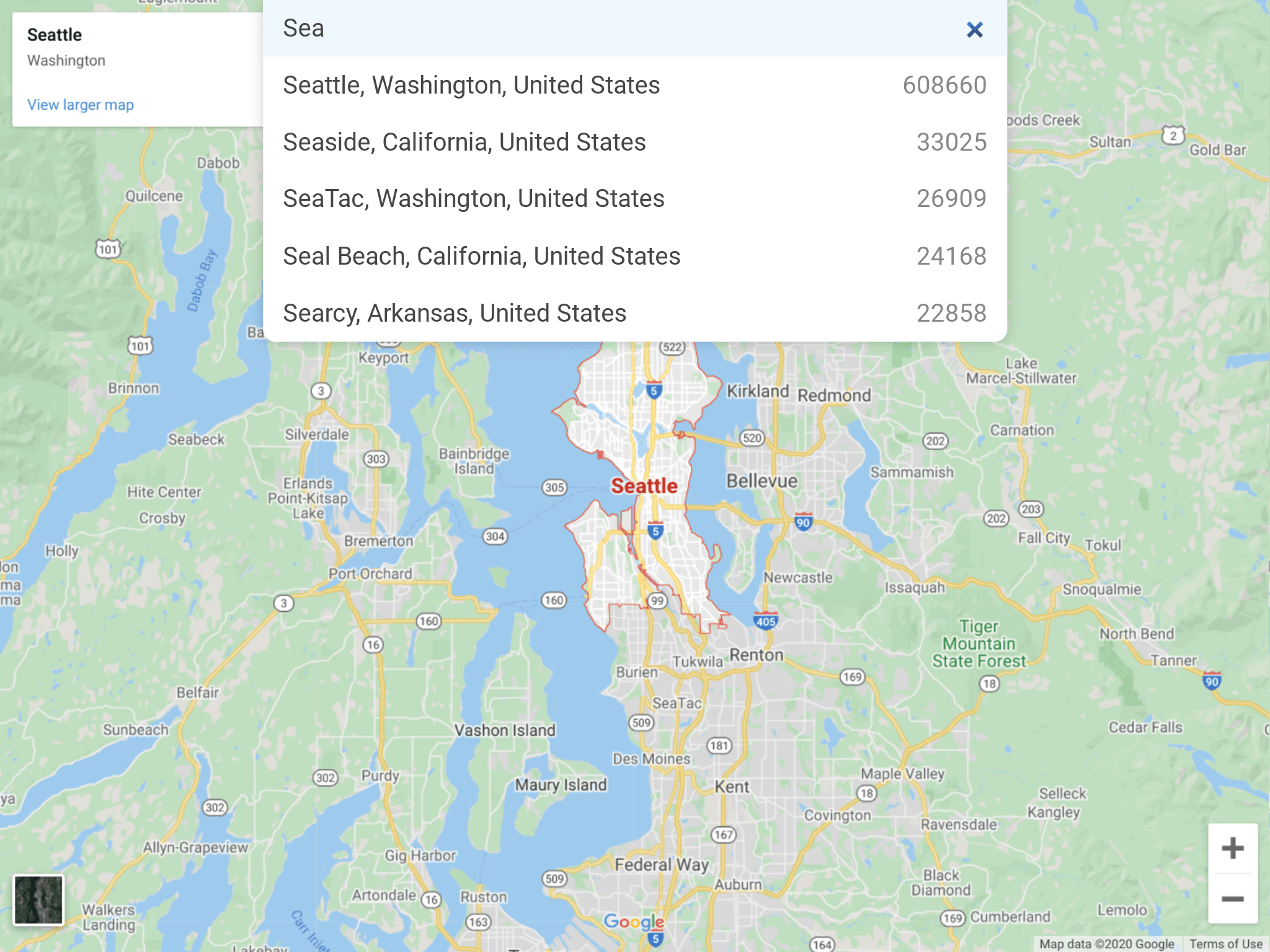}}%
  \hfill%
  \subfloat{\includegraphics[height=2.15in, valign=t]{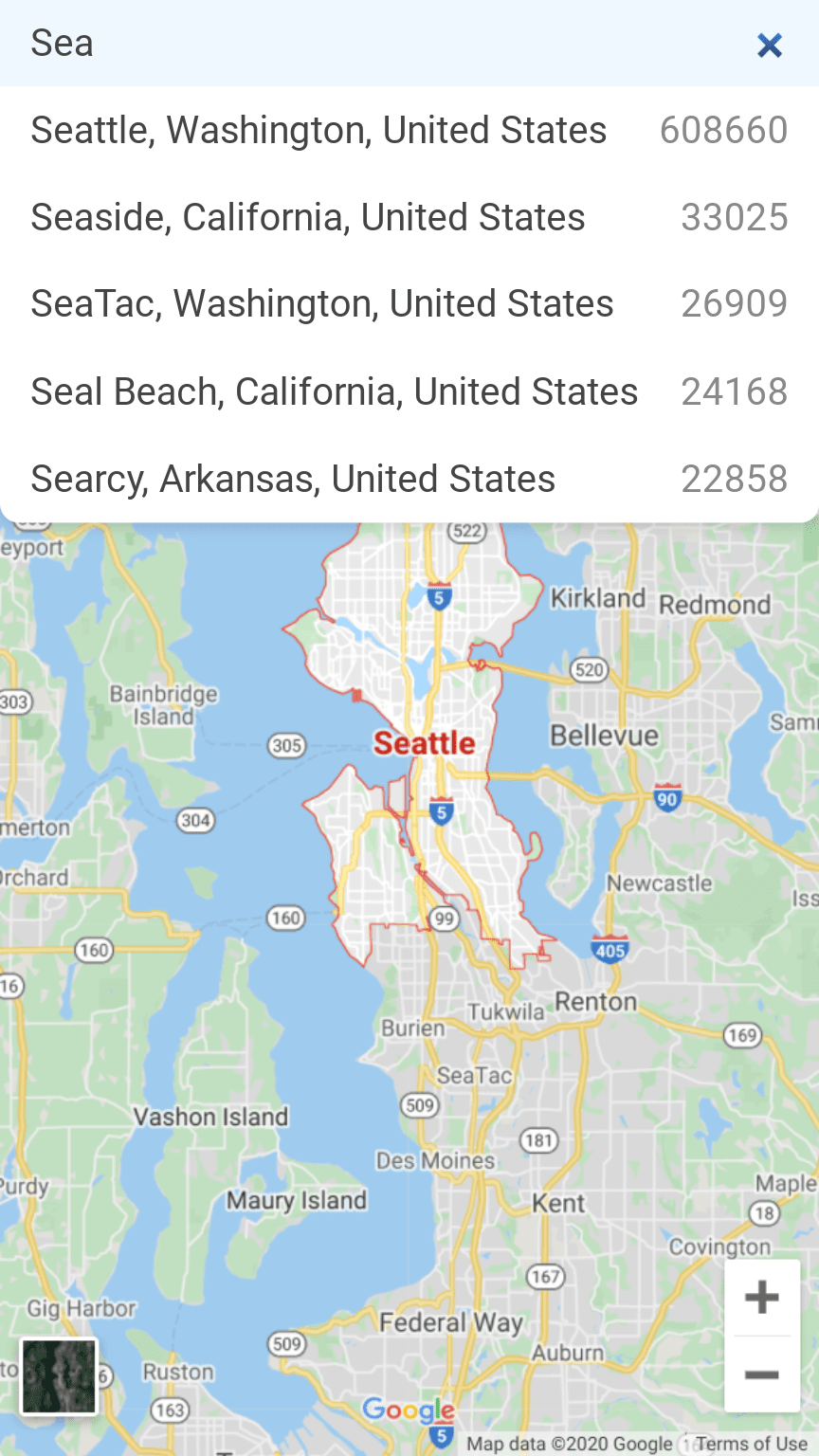}}%
  \caption{Text-based app compared to a simple web app for Autocomplete \cite{10.1145/2839509.2844678}.}
  \Description{From left to right: the console input/output for autocompleting city names given the query string "Sea" and two screenshots of the web app with the same query as it would appear on a tablet and on a smartphone.}
  \label{fig:teaser}
\end{teaserfigure}

\maketitle

\section*{Objective}
Socially relevant computing curricula that pursue sociocultural, political, and ethical connections to computing can broaden student interpretations of computer science's disciplinary values and support the personal identities that students bring to the classroom \cite{10.1145/2714569, 10.1080/07370008.2020.1730374}. One source of ideas for enabling these connections is the Nifty Assignments session, which has attracted computing educators at the annual ACM Technical Symposium on Computer Science Education (SIGCSE) since 1999. Over 120 assignments have been published across many areas of computing, such as algorithmic art and music, creative games, and interactive simulations \cite{nifty}. However, despite the materials provided, adopting and integrating new assignments can still be a time-consuming process for educators.

We present a tutorial to build web apps for any text-based programming assignment. Web apps can help educators connect existing assignments to their broader social contexts. However, the complexity of web technologies presents a steep barrier to entry. Rather than teach web technologies, the objective of this tutorial is to prepare educators with no prior web development experience to use, modify, and create \cite{10.1145/3304221.3319786} simple web apps based on predefined templates. Simple web apps take a text query as input and display a list of text results that automatically updates on input change. Simple web apps can be enhanced by binding app data to web services such as Google Maps (fig. \ref{fig:teaser}). By the end of the tutorial, participants will be able to create simple web apps for their own assignments in about one hour by adapting our open-source templates.\footnote{\url{https://kevinl.info/nifty-web-apps/}}

\section*{Outline}
\begin{description}[labelwidth=4em, align=right]
\item[ 5 min] Presenter introductions and web app demos.
\item[ 5 min] Web app introduction and activity walkthrough.
\item[ 5 min] Groups: Setup and group member introductions.
\item[10 min] Groups: Use the Letter Inventory web app.
\item[20 min] Groups: Modify the app for Random Sentence Generator.
\item[20 min] Groups: Create a simple Autocomplete app.
\item[10 min] Discussion board showcase and wrap-up.
\end{description}

The faculty first-author will lead the presentation while the undergraduate co-authors will assist participants during small-group work. Prior to starting this project, the presenters spanned a spectrum of comfort levels with respect to web app development, including some presenters that had no web app development experience. Each presenter thus provides unique expertise and perspective on developing web apps, which will help accommodate the variety of backgrounds that we expect participants will bring to the tutorial. Collectively, the presenters developed web apps for 7 programming assignments in a Java-based CS2 introductory programming course at an R1 university: 4 Nifty Assignments \cite{10.1145/2839509.2844678, 10.1145/1508865.1509031, 10.1145/299649.299809, 10.1145/3328778.3372574}, 2 assignments from other CS1 and CS2 courses \cite{bajillion, links}, and 2 original assignments.

This tutorial has been tested with a small group of educators with no web app development experience. While the tutorial is intended for educators interested in modernizing their text-based programming assignments, it is also designed to be accessible to undergraduate students with basic programming experience in either Java or Python. Prior to the symposium, the tutorial will receive further refinements based on our experience providing it to students in our CS2 course.

\section*{Expectations}
The goal of this tutorial is to prepare programming-focused CS educators to build simple web apps that respond to user interaction by running student-written assignment code. While the approach is not strictly limited to Python or Java programming languages (or introductory programming in general), it is easiest to get started in programming languages that provide basic HTTP web server implementations in their standard library. We expect participants to leave the tutorial ready to build simple web apps that dynamically respond to a text query by displaying a list of results.

We will begin the tutorial with a short walkthrough providing direct instruction on essential web app concepts and definitions. After the walkthrough, most of the time will involve participants working in groups of 3 to 6. In each group, participants will use, modify, and create \cite{10.1145/3304221.3319786} simple web apps based on our predefined templates. Instruction is designed for enabling asynchronous learning, including screenshots of important interface elements and subgoal labels for each task. At 10-minute intervals during the small-group work, presenters will check-in with each participant.

At the end of the tutorial, the faculty lead will highlight key points and collect participant feedback. It is important that we create a community of practice to support enhancing the participant experience beyond the 75-minute special session, so participants will be invited to share their work, continue conversations, and support future web app development in a private online discussion board.

\section*{Virtual Format}
The asynchronous design of the instructional materials supports late-comers as well as participants moving through the tutorial at different rates. The most challenging component to adapt will be groupwork. We will be particularly explicit about tutorial expectations at the beginning of the event so that participants know how to support each other within a group and escalate questions, especially if not every group has a presenter available for the entire tutorial. Presenters will explicitly check-in with individual participants to ensure every participant makes progress.

\section*{Suitability for a Special Session}
This special session is a tutorial. Its format will allow participants to follow a scaffolded and self-paced tutorial for developing simple web apps by adapting predefined templates for new problems while receiving support from each other as well as the presenters. The 75-minute duration is the right length for this tutorial since the purpose of the tutorial is not to teach web programming or web technologies in any serious depth. We instead focus on the practical skills that CS educators need to turn prototypical text-based apps into simple yet engaging, motivating, and realistic web apps.

\bibliographystyle{ACM-Reference-Format}
\bibliography{ms}

\end{document}